\documentclass{article}
\usepackage{spconf,amsmath,graphicx,hyperref}
\usepackage{amssymb}
\usepackage{xcolor}
\usepackage{bbm}
\usepackage{booktabs}
\usepackage{multirow}
\usepackage{nicematrix}
\usepackage[normalem]{ulem}
\usepackage{subcaption}

\renewcommand{\paragraph}[1]{\noindent\textbf{#1}\ }

\title{Improving Audio Question Answering with Variational Inference}
\name{Haolin Chen \thanks{This work received funding under NAST: Neural Architectures for Speech Technology, Swiss National Science Foundation grant \href{https://data.snf.ch/grants/grant/185010}{185010}.}}
\address{
  Idiap Research Institute, Martigny, Switzerland\\
  École Polytechnique Fédérale de Lausanne, Lausanne, Switzerland
}

\begin{document}
\ninept
\maketitle
\begin{abstract}
Variational inference (VI) provides a principled framework for estimating posterior distributions over model parameters, enabling explicit modeling of weight uncertainty during optimization. By capturing this uncertainty, VI improves the reliability of predictions, yielding better calibrated outputs. In this work, we investigate the benefits of VI for challenging multimodal understanding and reasoning by applying the Improved Variational Online Newton (IVON), a recent VI optimizer, to fine-tuning a multimodal large language model on audio question answering tasks. Our results show that VI not only enhances predictive accuracy but also significantly improves calibration, reducing the model's overconfidence. These advances further support risk-sensitive applications such as selective prediction, where reliable confidence estimates are crucial. 
\end{abstract}

\section{Introduction}
Audio question answering (AQA) \cite{dcase2025aqa,mmau,airbench,audiobench} is a challenging multimodal task that requires models to interpret complex acoustic scenes in response to natural language queries. Unlike audio classification or captioning, AQA requires the extraction of fine-grained acoustic cues, temporal dependencies, and contextual information from potentially long and intricate audio sequences, followed by the generation of precise, contextually grounded answers. Recent advances in large audio language models (LALMs) \cite{qwen2audio,salmonn,qwen2.5omni,audioflamingo3} have made substantial progress toward this goal by unifying audio representation learning and language reasoning within a single architecture. However, producing reliable predictions in multimodal understanding tasks such as AQA remains challenging. Models are often overconfident, poorly calibrated, and prone to errors they cannot recognize, particularly when fine-tuned with limited data. In such cases, predicted probabilities fail to reflect true confidence, undermining their reliability for risk-sensitive applications where robust uncertainty estimation is essential. While these challenges have been demonstrated in unimodal domains such as natural language processing \cite{yang2024bayesian,ivonlora}, we argue that they extend directly to multimodal scenarios.

Variational inference (VI) \cite{DBLP:conf/nips/Graves11,blundell2015weight,DBLP:conf/icml/KhanNTLGS18,DBLP:conf/nips/OsawaSKJETY19,ivon} provides a principled framework for addressing these limitations by modeling the uncertainty of neural network parameters. Unlike traditional deep learning optimizers that produce point estimates, VI models full posterior distributions over parameters, explicitly capturing weight uncertainty during training. This leads to more accurate predictive uncertainty estimation, better calibration, and improves risk-sensitive applications such as selective prediction (SP), where the model can abstain from answering when uncertain. Recent advances in VI \cite{ivon} demonstrate superior calibration and uncertainty quantification compared not only to conventional optimizers like Adam \cite{adam,adamw}, but also to uncertainty estimation methods such as Monte Carlo dropout \cite{mcdropout}, stochastic weight averaging \cite{swag}, and deep ensemble \cite{deepensemble}, while maintaining high efficiency and effectiveness for large-scale networks.

In this work, we utilize the Improved Variational Online Newton (IVON), a recent VI optimizer that matches or surpasses Adam’s accuracy with comparable computational cost, to fine-tune a state-of-the-art large audio language model on challenging audio question answering tasks. Our experiments show that IVON not only achieves superior predictive accuracy compared to Adam, but also produces substantially better-calibrated and more reliable predictions. This improves the trustworthiness of multimodal understanding and reasoning systems in risk-sensitive applications like selective prediction \cite{car,auc,DBLP:conf/eccv/WhiteheadPS0DRR22} where the reliable confidence scores are required.

\section{Audio Question Answering}
\subsection{Overview}
Audio question answering \cite{dcase2025aqa,mmau,airbench,audiobench} is an emerging multimodal challenge at the intersection of acoustic understanding and natural language processing. The task requires a system to listen to an audio clip, interpret its content, and generate a precise, contextually relevant answer to a question posed in natural language. Building upon the foundation of automated audio captioning, AQA extends beyond simple description to enable more interactive understanding and reasoning of audio. Unlike conventional audio classification which focuses on identifying pre-defined sound events, AQA requires a higher level of semantic comprehension and complex reasoning. The complexity of the task is multi-layered: a model must not only process raw audio and identify acoustic information but also understand temporal relationships between sounds, infer implicit context, and provide a coherent response to multiple-choice or open-ended questions. This combination of auditory perception and linguistic reasoning across potentially long audio sequences makes AQA a highly challenging task and a valuable benchmark for evaluating advanced reasoning capabilities of audio language models.

\subsection{Large Audio Language Models}
Recent advances in LALMs \cite{qwen2audio,salmonn,qwen2.5omni,audioflamingo3} are transforming audio processing by combining the reasoning capabilities of large language models (LLMs) with auditory understanding. Unlike traditional cascaded pipelines with separate modules for speech recognition and language processing, LALMs handle diverse audio inputs, including speech, music, and general sounds, within a single, unified architecture. A typical design uses a powerful audio encoder, often initialized from pre-trained models like Whisper \cite{whisper}, to convert raw audio into feature embeddings. These are then aligned with an LLM backbone, enabling audio-grounded reasoning and multi-turn dialogue. For example, Qwen2-Audio \cite{qwen2audio} and Qwen2.5-Omni \cite{qwen2.5omni} employ a Whisper-based encoder with a pooling layer to shorten audio representations. Similarly, Audio Flamingo 3 \cite{audioflamingo3} uses a unified AF-Whisper encoder trained across varied audio modalities. These techniques allow LALMs to move beyond simple transcription to tackle complex reasoning and problem-solving in audio intelligence.

\subsection{Selective Prediction}
Question answering (QA) is traditionally formulated as a prediction task in which the model must always provide an answer from a fixed answer space $\mathcal{A}$, regardless of uncertainty. Formally, this is represented as a function $f: \mathcal{X} \rightarrow \mathcal{A}$, where each input $x = (c, q) \in \mathcal{X}$ consists of the context $c$ (in AQA, the audio) and a corresponding question $q$. This formulation compels the model to produce an answer for every input, even in cases where it is uncertain or likely to be incorrect, potentially leading to unreliable predictions. To mitigate this issue, selective prediction (SP) allows the model to abstain from answering when appropriate, effectively enabling it to respond with ``I don’t know''.

Following the notation of \cite{DBLP:conf/eccv/WhiteheadPS0DRR22}, the selective QA task can be defined as a selective model $h: \mathcal{X} \rightarrow \mathcal{A} \cup \{ \emptyset \}$, where the model can either output an answer from $\mathcal{A}$ or choose to abstain (denoted by $\emptyset$). A typical approach is to construct $ h $ from two components: an answer predictor $ f $, and a confidence function $ g: \mathcal{X} \rightarrow [0,1] $, which estimates the reliability of the prediction. The model answers only when its confidence exceeds a threshold $ \gamma $; otherwise, it abstains:
\begin{equation}\label{eq:select}
h(x) = (f,g)(x) =
\begin{cases}
f(x) & \text{if } g(x) \geq \gamma, \\
\emptyset & \text{if } g(x) < \gamma.
\end{cases}
\end{equation}
The threshold $ \gamma $ governs the trade-off between coverage and reliability. A higher value leads to more conservative behavior, as the model answers only when highly confident. Lowering $ \gamma $ increases the number of answered questions but may reduce prediction accuracy. Ideally, $ g(x) $ should produce high values for correct predictions and low values for incorrect ones. The simplest confidence function is MaxProb, which utilizes the softmax probability of the model’s predicted class with the highest confidence. Given the output distribution $ p_i $ produced by a deterministic model, the selective function is defined as:
\begin{equation}\label{eq:g_maxprob}
    g_{MaxProb}(x) = p_k, \quad \text{where } k = \underset{i}{\operatorname{argmax}}\, p_i.
\end{equation}
We use MaxProb in our experiments as it can be applied to any deterministic model, regardless of calibration-enhancing techniques.

\begin{table*}[tbp]
  \centering
  \caption{Main results. The best results are marked in \textbf{bold}. ECE, Brier score, and AUC are $100\times$.}
    \begin{tabular}{llcccccccc}
    \toprule
    \multirow{2}[2]{*}{\textbf{Domain}\vspace{1ex}} & \multirow{2}[2]{*}{\textbf{Method}\vspace{1ex}} & \multirow{2}[2]{*}{\textbf{ACC}\,$\uparrow$\vspace{1ex}} & \multicolumn{3}{c}{\textbf{Calibration}} & \multicolumn{4}{c}{\textbf{Selective Prediction}} \\
          &       &       & \textbf{ECE}\,$\downarrow$ & \textbf{NLL}\,$\downarrow$ & \textbf{Brier}\,$\downarrow$ & \textbf{C@1\%}\,$\uparrow$ & \textbf{C@5\%}\,$\uparrow$ & \textbf{C@10\%}\,$\uparrow$ & \textbf{AUC}\,$\downarrow$ \\
    \midrule
    \multirow{3}[2]{*}{\textbf{BQA}\vspace{1ex}} & AdamW & 88.57  & 9.7   & 0.52  & 20.5  & 51.5  & 82.9  & 96.3  & 2.2  \\
          & IVON Mean & \textbf{89.02} & 7.4   & 0.39  & 17.4  & 61.8  & \textbf{85.8} & \textbf{98.4} & \textbf{1.8} \\
          & IVON MC-8 & 88.93  & \textbf{6.6} & \textbf{0.36} & \textbf{16.9} & \textbf{66.8} & 85.4  & 97.8  & \textbf{1.8} \\
    \midrule
    \multirow{3}[2]{*}{\textbf{TSQA}\vspace{1ex}} & AdamW & \textbf{67.39} & 26.2  & 1.42  & 57.0  & 4.2   & 6.3   & 29.9  & 16.8  \\
          & IVON Mean & 67.16  & 18.6  & 1.09  & 50.0  & 5.6   & 19.0  & 34.0  & 15.6  \\
          & IVON MC-8 & 67.16  & \textbf{15.6} & \textbf{0.99} & \textbf{47.7} & \textbf{6.1} & \textbf{24.6} & \textbf{36.0} & \textbf{15.1} \\
    \midrule
    \multirow{3}[2]{*}{\textbf{CQA}\vspace{1ex}} & AdamW & 84.21  & 12.7  & 0.71  & 28.0  & 1.4   & 51.4  & 84.2  & 6.2  \\
          & IVON Mean & \textbf{85.02} & 9.1   & 0.55  & 24.3  & 17.9  & 66.1  & 86.7  & 4.7  \\
          & IVON MC-8 & \textbf{85.02} & \textbf{7.9} & \textbf{0.51} & \textbf{23.5} & \textbf{19.6} & \textbf{67.1} & \textbf{87.0} & \textbf{4.5} \\
    \midrule
    \multirow{3}[2]{*}{\textbf{Domain Avg.}\vspace{1ex}} & AdamW & 80.06  & 16.2  & 0.88  & 35.1  & 19.0  & 46.8  & 70.2  & 8.4  \\
          & IVON Mean & \textbf{80.40} & 11.7  & 0.68  & 30.6  & 28.4  & 57.0  & 73.0  & 7.4  \\
          & IVON MC-8 & 80.37  & \textbf{10.0} & \textbf{0.62} & \textbf{29.4} & \textbf{30.8} & \textbf{59.0} & \textbf{73.6} & \textbf{7.2} \\
    \midrule
    \multirow{3}[2]{*}{\textbf{Weighted Avg.}\vspace{1ex}} & AdamW & 80.45  & 15.7  & 0.87  & 34.5  & 3.8   & 41.8  & 73.8  & 7.4  \\
          & IVON Mean & \textbf{80.97} & 11.2  & 0.67  & 30.1  & 16.6  & 56.2  & 76.6  & 6.0  \\
          & IVON MC-8 & \textbf{80.97} & \textbf{9.5} & \textbf{0.61} & \textbf{28.9} & \textbf{19.5} & \textbf{58.4} & \textbf{77.3} & \textbf{5.8} \\
    \bottomrule
    \end{tabular}%
  \label{tab:main_result}%
\end{table*}%

\section{Variational Inference}
\subsection{Overview}
Variational inference frames the task of modeling distributions over neural network parameters as an optimization problem. It approximates the intractable true posterior with a tractable surrogate distribution (typically Gaussian) and minimizes the Kullback–Leibler (KL) divergence between them. The evidence lower bound (ELBO) serves as the variational objective, enabling the use of stochastic gradient descent in modern deep learning frameworks for scalable and efficient training of probabilistic models. This makes VI a practical approach for training Bayesian neural networks \cite{DBLP:conf/icml/KhanNTLGS18,DBLP:conf/nips/OsawaSKJETY19,ivon}, where explicitly modeling parameter uncertainty can enhance generalization and robustness. Compared to conventional optimizers such as Adam \cite{adam,adamw}, VI offers additional benefits including improved predictive uncertainty estimation, better calibration, and the potential for model merging to support knowledge transfer.

In contrast to traditional deep learning approaches which estimate parameters by directly minimizing the empirical risk $\ell(\boldsymbol{\theta})$ via gradient descent, variational methods estimate a posterior distribution $q(\boldsymbol{\theta})$ over parameters by minimizing the following objective:
\begin{equation}
    \mathcal{L}(q) = \mathbb{E}_{q(\boldsymbol{\theta})}[\ell(\boldsymbol{\theta})] + \mathbb{D}_{KL}(q(\boldsymbol{\theta})||p(\boldsymbol{\theta}))
\end{equation}
where $p(\boldsymbol{\theta})$ is a prior distribution over parameters. The estimated posterior $q(\boldsymbol{\theta})$ enables sampling model parameters and averaging output logits across multiple Monte Carlo (MC) samples during inference, yielding better calibrated predictions. The optimization of $\mathcal{L}(q)$ is fundamentally different from minimizing $\ell(\boldsymbol{\theta})$ using gradient descent: it requires sampling from $q(\boldsymbol{\theta})$ during training and typically introduces twice as many parameters when using a Gaussian approximation with diagonal covariance. Early efforts in VI \cite{DBLP:conf/nips/Graves11,blundell2015weight} relied on stochastic gradient estimators to optimize the mean and variance parameters of $q(\boldsymbol{\theta})$, but these approaches did not scale well to modern architectures. More recent methods based on natural gradients \cite{DBLP:conf/icml/KhanNTLGS18,DBLP:conf/nips/OsawaSKJETY19} have shown promising results through Adam-like updates. However, they still underperform Adam and incur substantially higher computational costs.

\subsection{Improved Variational Online Newton (IVON)}
The Improved Variational Online Newton (IVON) \cite{ivon} is a recent VI optimizer that achieves performance matching or surpassing Adam while maintaining a comparable computational cost. 
Its key innovations include avoiding the costly per-example gradient-square computation via a reparameterization trick and integrating several practical techniques to enhance performance. 
Specifically, IVON models a diagonal Gaussian posterior $q(\boldsymbol{\theta}) = \mathcal{N}(\mathbf{m}, 1/\lambda(\mathbf{h} + \delta))$, where $\mathbf{m}$ is the mean equivalent to $\boldsymbol{\theta}$ in Adam, $\lambda$ is the effective sample size, $\mathbf{h} = \nabla^2\ell(\boldsymbol{\theta})$ is the diagonal Hessian, $\delta$ is the weight decay, while using a zero-mean isotropic Gaussian prior $p(\boldsymbol{\theta})$ with a scalar variance. 
Switching to IVON only requires a few lines of code change, making it a drop-in replacement for Adam. 
IVON has been shown effective when applied to fine-tuning language models with the low-rank adaptation (LoRA) \cite{lora} on natural language understanding and reasoning tasks for improving parameter efficiency and calibration \cite{vilora,ivonlora}.
An alternative and effective approach for predictive uncertainty quantification is the Laplace approximation \cite{yang2024bayesian}. However, it requires an extra pass through the data to compute the Hessian, resulting in significantly higher computational and memory costs, as well as increased implementation complexity.

Although IVON offers an Adam-like optimization framework with only about 1\% computational overhead from the sampling before the forward pass, some of its hyperparameters require careful configuration. Below, we highlight the key hyperparameters that could impact the training stability and the final performance.
\begin{enumerate}
    \item \textbf{Hessian initialization} ($h_0$): $h_0$ is typically in the range of $10^{-3}$ to $1$. A larger $h_0$ corresponds to a smaller initial variance and a more concentrated posterior. It can be set according to the scale of the gradient for the Hessian estimate to converge faster. A small $h_0$ can be used together with a large $\lambda$ to stabilize training.
    
    \item \textbf{Effective sample size} ($\lambda$): $\lambda$ scales the estimated variance and governs the amount of stochasticity introduced via sampling. A smaller $\lambda$ can lead to higher variance and potential training instability. In practice, $\lambda$ is often set to the size of the training dataset. For very small datasets, using a much larger $\lambda$ can help stabilize training and improve performance.
\end{enumerate}

\section{Experiments}
\subsection{Experimental Details}
\paragraph{Model.} We utilize the Qwen2.5-Omni 3B model in our experiments. Qwen2.5-Omni \cite{qwen2.5omni} is a state-of-the-art end-to-end multimodal large language model designed to perceive diverse modalities, including text, images, audio, and video, while simultaneously generating text and natural speech responses in a streaming manner. Given the vision encoder and the speech decoder are not useful for AQA tasks, we only fine-tune the thinker model with the low-rank adaptation (LoRA) \cite{lora} while keeping the audio encoder frozen. LoRA is applied to all linear layers, with a rank of 8 and an alpha of 16, without applying dropout.

\paragraph{Data.} 
The DCASE 2025 AQA dataset \cite{dcase2025aqa} is used in our experiments, which comprises three multiple-choice question answering subsets designed to evaluate distinct aspects of audio-language understanding and reasoning. (1) Bioacoustics QA (BQA) focuses on fine-grained auditory grounding in the bioacoustic domain, requiring models to identify marine mammal vocalizations (31 species) and reason about their acoustic traits and ecological context. It contains 0.7k/0.2k train/dev QA pairs with diverse sampling rates and durations. (2) Temporal Soundscapes QA (TSQA) consists of 1k/0.6k pairs that aim to assess temporal reasoning capabilities by presenting models with questions concerning the classification and temporal structure of overlapping or sequential sound events, with questions on event ordering, onset/offset detection, and duration estimation in 10s clips. (3) Complex QA (CQA) includes 6.4k/1.6k pairs covering real-world auditory scenarios that challenge models with multi-faceted questions that require integration of temporal, acoustic, and contextual cues to interpret overlapping events, auditory sequences, and abstract relational patterns.

\paragraph{Training and evaluation.} All models are trained for 3 epochs with a batch size of 4 on the full training set. For IVON, we use an effective sample size $\lambda$ of $10^7$, a Hessian initialization $h_0$ of $1\times10^{-3}$, a learning rate of $0.03$ with cosine learning rate decay to 0, and a weight decay $\delta$ of 0. For Adam, we train the model using the AdamW optimizer without weight decay and a learning rate of $5\times10^{-5}$ with cosine learning rate decay to 0. The model is provided with the audio sequence followed by a question and several options (e.g., ``A. ..., B. ...''); the task is to predict the correct option (e.g., ``A''). Evaluation is performed on the development set with results reported on three domains respectively. In addition to the averaged results across all samples (Weighted Avg.), the averaged scores across three domains (Domain Avg.) are also calculated. For IVON, results are reported under two settings: predictions evaluated at the posterior mean (IVON Mean), and the average over 8 Monte Carlo samples (IVON MC-8). All results are averaged over 10 runs with different random seeds.

\subsection{Evaluation Metrics}
\paragraph{Calibration.}
To evaluate how well the model’s predicted confidences align with its actual correctness, we report three complementary calibration metrics in addition to predictive accuracy (ACC). Expected calibration error (ECE) \cite{ece} measures the average discrepancy between predicted probabilities and empirical accuracies across different confidence levels. Negative log likelihood (NLL) assesses how much probability the model assigns to the correct labels, penalizing low-confidence predictions for the true class. Brier score captures the mean squared difference between predicted probabilities and actual labels, with lower scores indicating more accurate and better-calibrated probability estimates.

\paragraph{Selective prediction.}
We assess the model’s ability to abstain from answering when uncertain using two metrics. Coverage at risk (C@R) \cite{car} quantifies the maximum proportion of inputs the model can answer while keeping the error rate below a specified threshold, reflecting its effectiveness in maintaining reliability under a risk constraint. Area under the risk-coverage curve (AUC) \cite{auc} summarizes performance across all confidence thresholds by integrating the risk-coverage curve; lower values indicate the model maintains lower risk over a wider range of coverages.

\begin{figure}[tbp]
\centering
\resizebox{\columnwidth}{!}{
\includegraphics[width=\columnwidth]{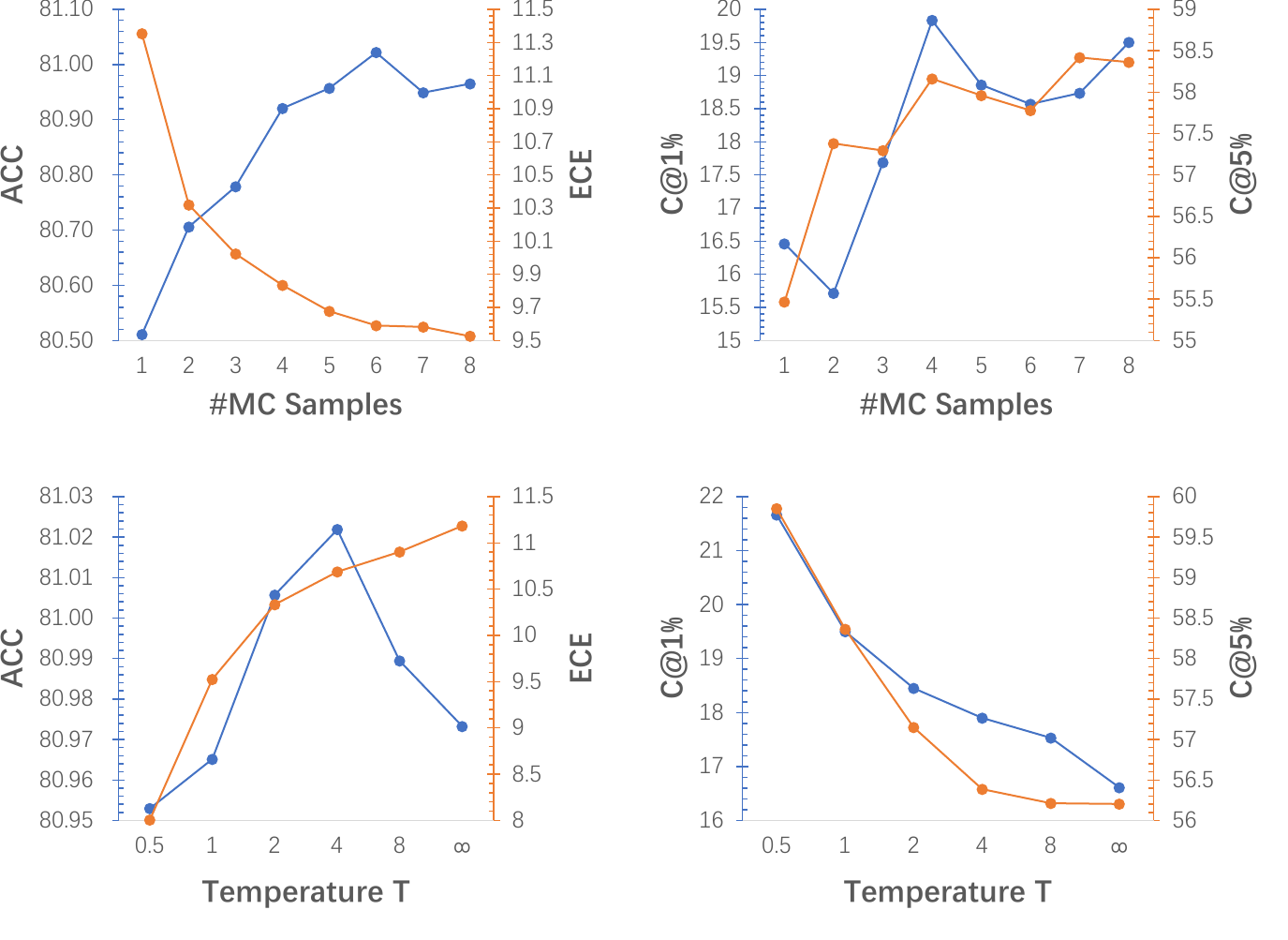}
}
\caption{Performance with varying MC samples and temperature.}
\label{fig:varying}
\end{figure}

\section{Results and Analyses}
The main results are shown in Table \ref{tab:main_result}. We elaborate our findings from the following perspectives.

\subsection{Accuracy}
In terms of predictive accuracy, models fine-tuned with IVON achieve performance comparable to or slightly exceeding that of the AdamW baseline. On the two smaller subsets, BQA and TSQA, the results are nearly identical—IVON leads by roughly one example on BQA, while AdamW holds the same margin on TSQA. On the more challenging CQA task, both IVON Mean and IVON MC-8 outperform the baseline, reaching 85.02\% accuracy compared to 84.21\% for AdamW. Across datasets, IVON Mean and IVON MC-8 typically yield similar accuracy, suggesting that eight Monte Carlo samples are sufficient to match posterior-mean performance. When considering the weighted average over all tasks, both IVON variants achieve 80.97\%, surpassing AdamW’s 80.45\%. For reference, directly prompting the base model yields a domain average of 53.8\%. These results indicate that incorporating variational inference generally provides an improvement in predictive accuracy.

\subsection{Calibration}
The results clearly show that variational inference produces substantial calibration improvements. Across all datasets and metrics, the IVON models consistently outperform the AdamW baseline. For instance, on the CQA task, AdamW yields an ECE of 12.7, which is reduced to 9.1 by IVON Mean and further to 7.9 by IVON MC-8. This trend holds for NLL and Brier scores as well. Notably, the IVON MC-8 method, which leverages multiple samples from the learned posterior distribution to make predictions, consistently achieves the best calibration. Considering the domain average, IVON MC-8 reduces the ECE from 16.2 (AdamW) to 10.0 and the NLL from 0.88 to 0.62. This demonstrates that explicitly accounting for parameter uncertainty during inference through Monte Carlo sampling provides a more reliable and less overconfident model.

\subsection{Selective Prediction}
Models trained with IVON demonstrate substantially stronger performance in selective prediction, with significantly improved ability to identify and abstain from low-confidence predictions. For example, on the CQA dataset, when abstaining from just 1\% of the most uncertain predictions (C@1\%), the AdamW model's coverage is a mere 1.4\%, whereas IVON MC-8 achieves a coverage of 19.6\%. This indicates that the predictions of models trained by IVON are most confident about are far more likely to be correct. This pattern is consistent across all rejection fractions and datasets. The weighted average results show that IVON MC-8 improves the C@5\% coverage from 41.8\% to 58.4\% and reduces the overall selective prediction AUC from 7.4 to 5.8. These results confirm that the uncertainty estimates derived from the variational approach are not only better calibrated but also more effective for practical decision-making, allowing the model to reliably identify its own potential failures.

\subsection{Varying Sampling Configurations}
To assess the impact of sampling configurations during inference, we analyze how performance (weighted average) varies with the number of Monte Carlo (MC) samples and with the scaling of the posterior variance, as shown in Figure \ref{fig:varying}.

\paragraph{Number of MC samples.}
Increasing the number of MC samples generally improves both accuracy and calibration. ECE decreases consistently as more samples are used, though the gains diminish beyond a certain point. Coverage-at-risk metrics follow a similar trend but with greater variability. Using very few samples ($\leq4$) can yield lower accuracy and coverage than simply using the posterior mean without sampling, suggesting that sampling is beneficial only when a sufficient number of MC samples are employed.

\paragraph{Scale of variance.}
The effective sample size $\lambda$ can be rescaled at inference to control the amount of noise injected into model samples. We parameterize this using a temperature $T$ such that $\lambda_{\text{infer}} = T\lambda$, where a larger $T$ corresponds to a more concentrated posterior; $T \to \infty$ recovers the posterior mean (IVON Mean). As $T$ increases, accuracy improves slightly, but calibration deteriorates. Conversely, modestly reducing $T$ can enhance calibration with only a small accuracy loss. However, we notice that reducing $T$ too far destabilizes predictions, leading to sharp declines in both accuracy and calibration. This instability may be mitigated by increasing the number of MC samples, albeit at the cost of additional computation.

\section{Conclusions}
In this work, we employed the Improved Variational Online Newton (IVON), a recent variational inference optimizer, to fine-tune a multimodal large language model on challenging audio question answering tasks. Our results show that applying variational inference not only improves predictive accuracy but also produces better-calibrated predictions, mitigating overconfidence. The resulting improvements in calibration are especially valuable in risk-sensitive settings, such as selective prediction, where reliable confidence measures are essential. A current limitation is that our evaluation was restricted to single next-token prediction tasks. Future work includes investigating whether variational inference can benefit free-form generation tasks, such as open-ended question answering and speech synthesis.

\pagebreak

\bibliographystyle{IEEEbib}
\bibliography{chl-ref-short}

\end{document}